\documentclass[%
 reprint,
superscriptaddress,
 amsmath,amssymb,
 aps,
prb,
]{revtex4-2}

\usepackage{graphicx}
\usepackage{dcolumn}
\usepackage{bm}
\usepackage{hyperref}

\usepackage{physics}
\usepackage{graphicx}
\usepackage{amsmath}
\usepackage{amssymb}
\usepackage{amsfonts}
\usepackage{color}
\usepackage{xspace}
\usepackage{subfigure}
\usepackage{bm}
\usepackage{bbold}
\usepackage{blkarray}
\usepackage{float}
\usepackage[normalem]{ulem}
\usepackage{natbib}
\usepackage{cancel}
\usepackage{cleveref}

\setcitestyle{super}

\DeclareRobustCommand*{\citen}[1]{%
  \begingroup
    \romannumeral-`\x 
    \setcitestyle{numbers}%
    \cite{#1}%
  \endgroup   
}

\begin{document}
	

\preprint{APS/123-QED}

%

\title{Raman scattering signatures of spinons and triplons in frustrated antiferromagnets}

\author{O. R. Bellwood}
\email{olibellwood@gmail.com}
\affiliation{School of Mathematics and Physics, The University of Queensland, Brisbane, Queensland, 4072, Australia}

\author{H. L. Nourse}
\affiliation{Quantum Information Science and Technology Unit, Okinawa Institute of Science and Technology Graduate University, Onna-son, Okinawa 904-0495, Japan}

\author{B. J. Powell}
\email{powell@physics.uq.edu.au}
\affiliation{School of Mathematics and Physics, The University of Queensland, Brisbane, Queensland, 4072, Australia}

\graphicspath{ {./figs/} }

%

\begin{abstract}
Magnetically frustrated spin systems compose a significant proportion of topological quantum spin liquid candidates. Evidence for spin liquids in these materials comes largely from the detection of fractionalised spin-$1/2$ quasiparticles, known as spinons. However, the one-dimensional Heisenberg chain, which is topologically trivial, also hosts spinons. Thus, observing spinons does not necessarily signify long-range entanglement. 
Here, we show that spinons arising from one-dimensional physics leave a clear fingerprint in magnetic Raman scattering. We achieve this by calculating the magnetic Raman intensity of coupled Heisenberg chains. Our findings are in excellent agreement with the magnetic Raman scattering measurements on the anisotropic triangular antiferromagnet Ca$_3$ReO$_5$Cl$_2$.
\end{abstract}

\maketitle


There remains no unambiguous experimental test for a quantum spin liquid (QSL)\cite{anderson1973resonating, savary2016quantum, zhou2017quantum}. The most intuitive definition of a QSL, the absence of magnetic order in the groundstate of an interacting spin system\cite{balents2010spin}, would require us to demonstrate the absence of a property at a temperature (0~K) that experiment cannot access. Furthermore, this definition admits a number of other phases that  do not meet the stricter, modern criteria: the presence of topological order and fractional spin excitations \cite{knolle2019field}. For example, the groundstate of the one-dimensional $S=1/2$ Heisenberg antiferromagnet (Heisenberg chain)\cite{bethe1931theorie, MullerBB} is quantum disordered. And, although fractionalised, the spin-$1/2$ quasiparticles, known as spinons,  are topologically trivial as they cannot be braided \cite{broholm2020quantum}. Nevertheless, the experimental search for quantum spin liquids  largely rests on the detection of spinons. Therefore, one must be cautious of the dimensional origins of the excitations. Experimental signatures of spinons are only  evidence for a quantum spin liquid  if the material  is demonstrably not quasi-one-dimensional (q1D) \cite{lake2005quantum}. 

Here, we present a theory of inelastic light scattering from spinon excitations in q1D antiferromagnets and show that magnetic Raman scattering provides clear signatures of q1D physics. We derive the magnetic Raman intensity of the Heisenberg antiferromagnet on the anisotropic triangular lattice by invoking a q1D treatment of exchange scattering  (often called two-magnon scattering -- although this is inappropriate for fractionalised quasiparticles), including an explicit treatment of the (anti)bound `triplon' excitations\cite{Kohno}. 

\begin{figure}
    \centering
\includegraphics[width=0.9\columnwidth]{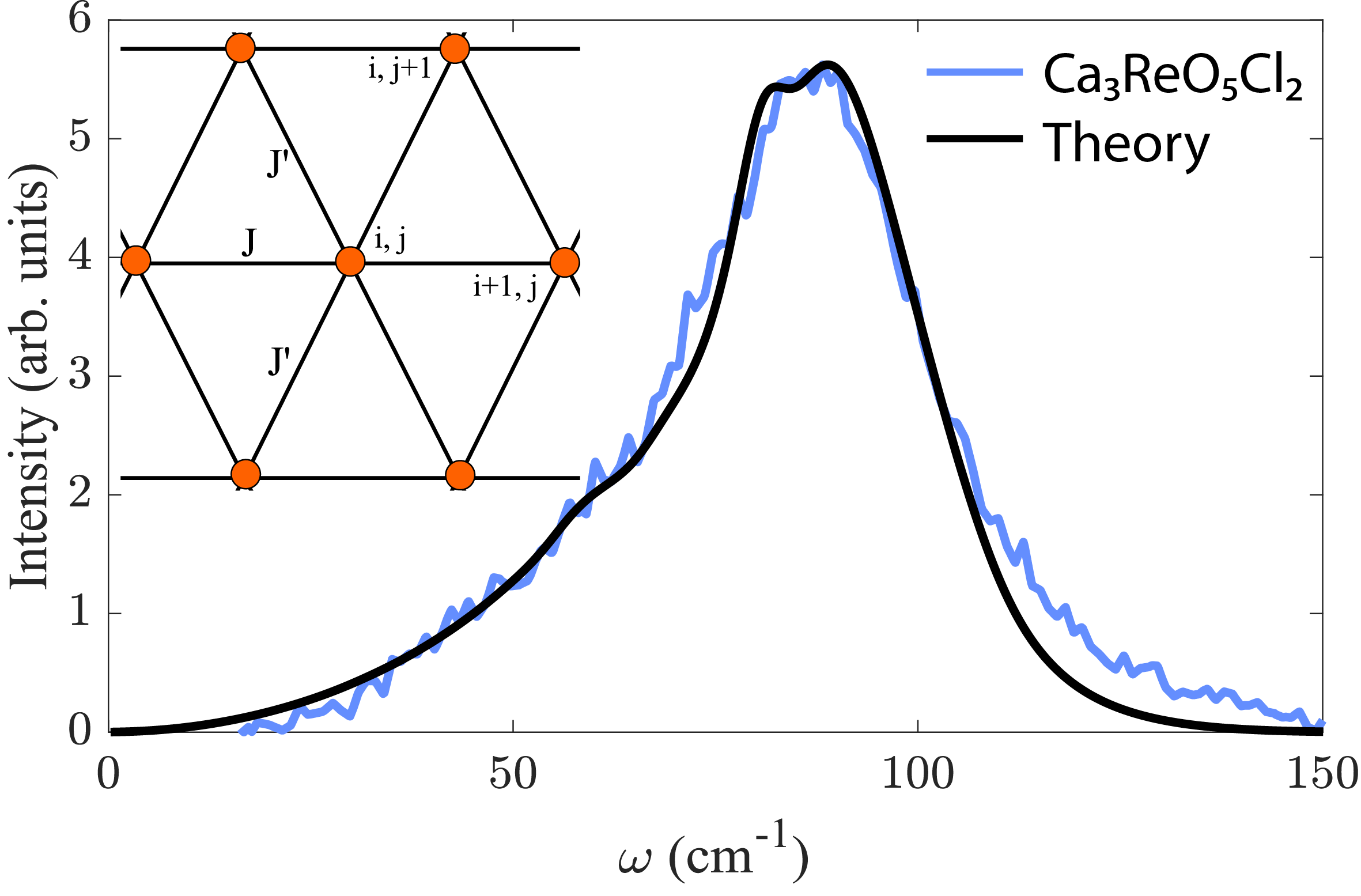}
\caption{\textbf{Parameter free prediction of the magnetic Raman intensity, $I(\omega)$, shows excellent agreement with the measured Raman scattering  from Ca$_3$ReO$_5$Cl$_2$ \citen{Choi}}. The black line is the calculated spectrum for the spin-1/2  anisotropic triangular lattice Heisenberg antiferromagnet (inset) with intrachain coupling $J = 26.7\text{cm}^{-1}$ and interchain coupling $J' = 0.3 J$   -- as previously estimated for Ca$_3$ReO$_5$Cl$_2$ \citen{hirai2019one,Choi}, and scattering polarisations $\theta = \phi = 0$. The blue line is the cross polarisation Raman data ($\theta = 0, \phi = 90^\circ$) subtracted from the parallel polarisation ($\theta = \phi = 0$) Raman scattering from Ca$_3$ReO$_5$Cl$_2$ measured at $4.3$~K. We predict zero intensity for the cross polarised magnetic Raman spectrum and remove it from the parallel polarised data as a potential  non-magnetic or non-intrinsic contribution. As the experiment was reported in arbitrary units only the lineshapes should be compared. The maxima of theoretical and experimental intensities have been matched to ease this comparison. }
\label{first_fig}
\end{figure}

We show that there are two key features in the magnetic Raman intensity of q1D antiferromagnets. (1) Two peaks appear either side of a Raman shift of $\omega = \pi J$, with a spliting $\sim J'$, where $J$ ($J'$) is the intrachain (interchain) coupling. The  double peak structure has a simple explanation. The lower energy peak arises from a van Hove singularity in the triplon dispersion. The higher energy peak is broader and is due to scattering from two-spinon excitations. (2) A $\cos^2(\theta + \phi)$ dependence on the polarization of the incoming $(\theta)$ and outgoing $(\phi)$ light relative to the chain axis. Together, these provide a robust indication of quasi-one-dimensionality, and by extension is an experimental `smoking gun' to rule out a topological quantum spin liquid phase. With exchange interactions previously estimated from experiment\cite{hirai2019one,Choi}, both the lineshape and polarization dependence of our theory quantitatively match recent magnetic Raman scattering measurements\cite{Choi} on the frustrated antiferromagnet Ca$_3$ReO$_5$Cl$_2$, Fig. \ref{first_fig}.

The dimensionality of  candidate quantum spin liquids can be difficult to determine. There must be a crossover between dimensionalities for anisotropic models\cite{Schulz,Troyer,PhysRevB.74.012407}. For example, the anisotropic triangular lattice, Fig. \ref{first_fig}(inset), is 1D for $J'/J\rightarrow0$, 2D for $J'=J$, and interpolates between these limits for $0 < J' < J$. Furthermore, it has been argued that `dimensional reduction'\cite{balents2010spin}  means that the ground state has a lower dimensionality than one would expect from the crystal structure, found in 
Cs$_2$CuCl$_4$ \cite{Kohno}, 
Ca$_3$ReO$_5$Cl$_2$ \cite{Choi,Zvyagin2022},
Ba$_4$Ir$_3$O$_10$ \cite{Cao2020},
BaCuSi$_2$O$_6$ \cite{Sebastian2006},
ZnCr$_2$O$_4$ \cite{ZnV2O4},
pharmacosiderite \cite{Okuma2021},
the {$X$[Pd(dmit)$_2$]$_2$} family \cite{Kenny1,Kenny2,Ido2022}, 
[(C$_3$H$_7$)$_3$NH]$_2$[Cu$_2$(C$_2$O$_4$)$_3$](H$_2$O)$_{2.2}$ \cite{PrattSR},  
{[(C$_2$H$_5$)$_3$NH]$_2$Cu$_2$\-(C$_2$O$_4$)$_3$}\cite{PrattJACS,Jacko,Dissanayake}, 
and in some models with Kitaev interactions \cite{Shannon,feng2023dimensional_nphys_fix}; implying that the dimensionality of a material can be emergent\cite{Sebastian2006}. Furthermore, dimensionality can be modified chemically and by applying pressure or stress.\cite{Zvyagin2019,Powell_RPP,Pustogow}
Thus, finding experimental methods to characterise the dimensionality of  quantum disordered phases of matter is imperative.

Inelastic neutron scattering is the gold standard experiment for probing quantum magnets as it provides direct access to  (the Fourier transform of) the spin-spin correlation function. 
However,  for many quantum spin liquid candidates it has proved impossible to grow the large single crystals required for inelastic neutron scattering.
Magnetic Raman scattering \cite{LF, cottam1986light, moriya1967theory, shastry1990theory, lemmens2003magnetic, devereaux2007inelastic} experiments can be performed on $\mu\text{m}^{3}$ sized sample volumes. Although magnetic Raman scattering is well established in ordered magnetic systems \cite{LF, cottam1986light}, the interpretation of experiments on quantum disordered antiferromagnets remains challenging\cite{wulferding2019raman,knolle2014raman,PatrickLee, PhysRevB.82.144412}. 

This motivates us to consider the antiferromagnetic Heisenberg model on the anisotropic triangular lattice, 
\begin{equation} \label{Ham_rec}
\begin{split}
\hat{H} & = \sum\limits_{i,j}  \left(  J \mathbf{S}_{i+1,j} + J'\mathbf{S}_{i,j+1} + J'\mathbf{S}_{i+1,j-1} \right) \cdot\mathbf{S}_{i,j}, \\
\end{split}
\end{equation}
where $\mathbf{S}_{i,j}$ is the  spin-1/2 operator at lattice site $(i,j)$, 
and $0 < J' < J$. This model can be reframed as an array of  Heisenberg chains, $\hat{H}_0 = J \sum_{i,j} \mathbf{S}_{i+1,j}\cdot\mathbf{S}_{i,j}$,  perturbed by the interchain interactions, $\hat{H}' = J'\sum_{i,j}  \left(\mathbf{S}_{i+1,j} +  \mathbf{S}_{i+1,j} \right)\cdot\mathbf{S}_{i,j}$. This framing is valid even in the regime of relatively large values of $J'$, as magnetic frustration quenches the spin correlations between chains up to  $J' \lesssim 0.7 J$ for the anisotropic triangular lattice \cite{PhysRevB.74.012407}.  

The exchange scattering Raman intensity is given by Fermi's Golden rule,
\begin{equation} \label{LF_vertex1}
\begin{split}
I(\omega, \theta, \phi) & = 2\pi \sum_{\lambda} \abs{\bra{\text{GS}}\hat{R}(\theta, \phi) \ket{\lambda}}^2 \delta(\omega - \omega_{\lambda}),  
\end{split}
\end{equation}
 where $\ket{\lambda}$ are the eigenstates of equation \eqref{Ham_rec} with eigenenergies $\omega_{\lambda}$, $\ket{\text{GS}}$ is the groundstate, and the Fleury-Loudon  operator is  \cite{LF}
\begin{equation} \label{Raman_Op1}
\hat{R}(\theta, \phi) = \sum_{i,j,\hat{\delta}} f_{\hat{\delta}}(\theta, \phi)  J_{\hat{\delta}} \,\, \mathbf{S}_{i,j} \cdot \mathbf{S}_{i+ \delta_{i}, j + \delta_{j}}.
\end{equation}
The sum over the lattice unit vectors, $\hat{\delta} = (\delta_i, \delta_j)$, picks out the relevant couplings, $J_{\hat{\delta}}$, and the polarisation dependent term  $f_{\hat{\delta}}(\theta, \phi) = (\hat{\varepsilon}_{\text{in}} \cdot \hat{\delta}) (\hat{\varepsilon}_{\text{out}} \cdot \hat{\delta})$ is defined with respect to the polarisation angles of the incoming and outgoing light relative to the $\hat{x}$-axis (parallel to the chains): $\varepsilon_{\text{in }}  = \cos \theta \hat{x} + \sin \theta \hat{y}$ and $\varepsilon_{\text{out}}  = \cos \phi \hat{x} + \sin \phi \hat{y}$ respectively. 

At first order in the interchain coupling the Raman matrix elements are
\begin{equation} \label{to_find_Gen}
\begin{split}
& \kern-2em \bra{\text{GS}}\hat{R}(\theta, \phi) \ket{\lambda} \\
& \approx \Big( \,\,_0\!\bra{\text{GS}} + \,\,_1\!\bra{\text{GS}} \Big) \hat{R}(\theta, \phi) \ket{\lambda} \\
& = \,_0\!\bra{\text{GS}} \hat{H}' \ket{\lambda} \sum_{\hat{\delta}'} \left[ f_{\hat{\delta}'}(\theta, \phi)- f_{\parallel}(\theta, \phi) \right] , 
\end{split}
\end{equation}
where the sum over $\hat{\delta}'$ runs over the interchain lattice vectors, $f_{\parallel}(\theta, \phi)$ is the intrachain polarisation factor for $\hat{\delta}_{\parallel} = (1,0)$,   $\ket{\text{GS}}_0$ is the unperturbed groundstate of equation (\ref{Ham_rec}) made entirely of groundstate Heisenberg chains\cite{bethe1931theorie},  $\ket{\text{GS}}_1 = (\omega_0 - \hat{H}_0)^{-1}\hat{H}'\ket{\text{GS}}_0$ is the first order perturbative correction to the groundstate in the interchain interaction, and $\omega_0$ is the unperturbed groundstate energy. 

For the anisotropic triangular lattice  $\sum_{\hat{\delta}'} \left[ f_{\hat{\delta}'}(\theta, \phi) - f_{\parallel}(\theta, \phi) \right] = -(3/2) \cos(\theta + \phi)$. Thus, for a q1D state  $I(\omega, \theta, \phi) \propto \cos^2(\theta + \phi)$. Hence, parallel polarisation  ($\theta = \phi = 0$) will give maximal intensity and cross polarisation  ($\theta = 0, \phi = 90^\circ$) will yield zero intensity. Henceforth we discuss only the parallel polarisation Raman intensity $I(\omega)\equiv I(\omega,0,0)$ as other polarisations can be trivially calculated from this. 

In contrast, on general symmetry grounds, one expects the Raman scattering from $C_{2h}$ symmetric systems (such as the anisotropic triangular lattice) to be of the form $I(\omega, \theta, \phi) = I_1(\omega) \cos^2(\theta+\phi) +  I_2(\omega) \sin^2(\theta+\phi)$. Both $I_1(\omega)$ and $I_2(\omega)$, which correspond to scattering in the $A_1$ and $A_2$ channels respectively, appear at leading order in the Fleury-Loudon vertex and therefore should have similar magnitudes. Thus, if the ground state retains the full symmetry of the Hamiltonian we generically expect that the Raman scattering will not depend strongly on the polarisation of the incoming or outgoing light. All of the explicit calculations we are aware of for Raman scattering from 2D quantum spin liquids find little\cite{knolle2014raman} to no\cite{PatrickLee,Cepas} dependence on  $\theta$ or $\phi$. Thus, $I(\omega, \theta, \phi) \propto \cos^2(\theta + \phi)$ in a quantum disordered system is a strong indication of q1D physics.

Two spinons can propagate between chains by binding to form a triplon, which may be either bound (below the energy of the spinon continuum) or antibound (above the spinon continuum)\cite{Kohno}. The states that contribute to the Raman intensity at first order contain two pairs of spinon excitations, analogous to a two-magnon state in the Fleury-Loudon formalism\cite{LF}. The spinon pairs exist on different chains with equal and opposite centre-of-mass momentum and total $S^z = 0$ in accordance with exchange Raman scattering selection rules\cite{LF}. Therefore, there are multiple classes of the excitations: 

\textit{Spinonic}: Neither pair of spinons forms a triplon. The energy of these states is largely goverened by the  two-spinon dispersion. 

\textit{One-triplon plus two-spinon}: One  spinon pair forms a triplon, which can be either bound or antibound. The other pair remains spinonic. 

\textit{Two-triplons} Both spinon pairs form triplons. Whether the triplons are bound or antibound is determined by the Fourier component of the interchain coupling\cite{Kohno}, and is the same for both  pairs. 

To evaluate the matrix element $_0\!\bra{\text{GS}} \hat{H}' \ket{\lambda}$  we approximate  $\ket{\lambda}$ by the basis of Heisenberg chain eigenstates where two different chains  contain a two-spinon excitation, and all other chains are the single chain groundstate. 
This allows us to construct an effective Schr\"odinger equation from equation (\ref{Ham_rec}) that we solve numerically, see Methods. 

\begin{figure}
    \centering
\includegraphics[width=0.9\columnwidth]{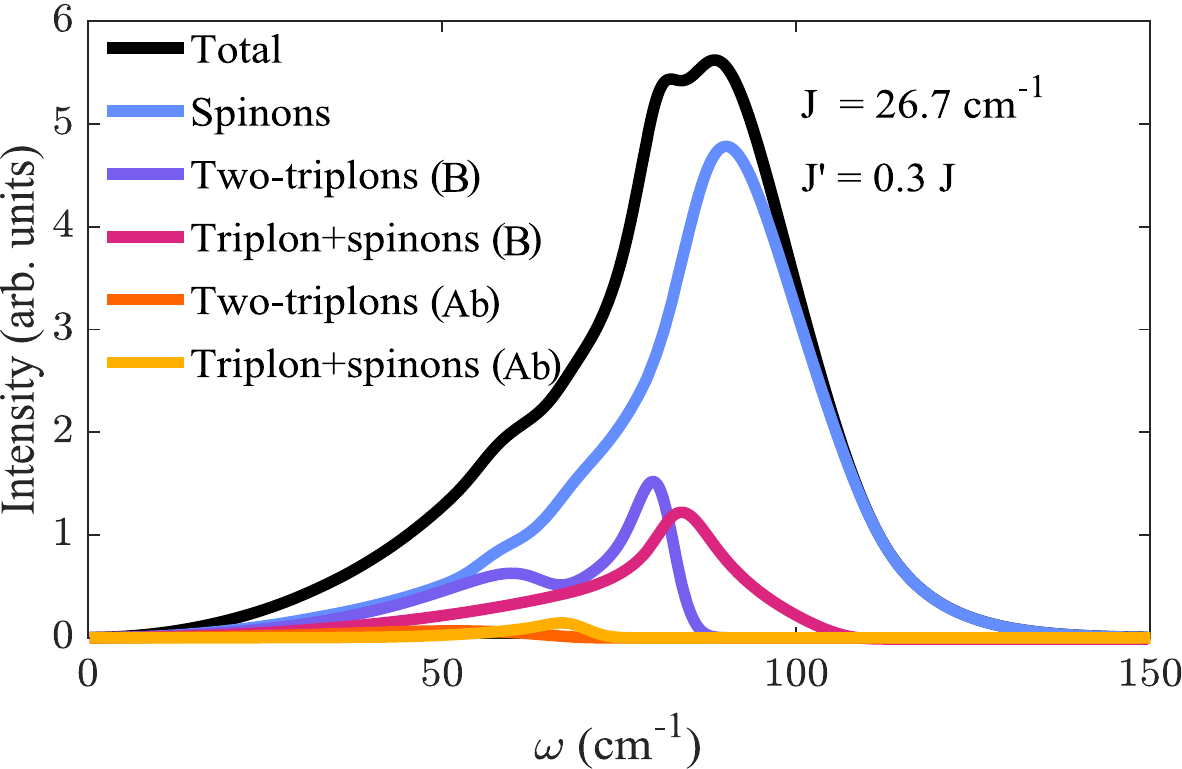}
\caption{\textbf{Decomposition of the magnetic Raman intensity of the anisotropic triangular lattice into excitation classes.} Triplons can be either bound (B) or  antibound (Ab). The proximity of the bound two-triplon, bound-triplon plus two-spinon, and the spinonic intensity peaks leads to a  shoulder and a flat-topped peak around a Raman shift of approximately $\omega = 85\text{cm}^{-1}$ in the total intensity. Chosen parameters are for Ca$_3$ReO$_5$Cl$_2$.}
\label{second_fig}
\end{figure}

The quantum disordered state in Cs$_2$CuCl$_4$ is an obvious test case for our theory, as the neutron scattering is well understood in terms of the anisotropic triangular lattice in the q1D regime.\cite{Kohno}
However, magnetic Raman scattering has not yet been reported for Cs$_2$CuCl$_4$.
The inelastic neutron scattering  spectrum of Ca$_3$ReO$_5$Cl$_2$ is remarkably similar to that of Cs$_2$CuCl$_4$ and shows clear evidence for spinon and triplon excitations\cite{hirai2017visible, hirai2019one, nawa2020bound}. Furthermore, both materials are believed to be described by the Heisenberg model on the anisotropic triangular lattice with $J'\simeq 0.3-0.35 J$ \cite{hirai2019one,Choi,coldea2002direct}. Magnetic Raman scattering was recently reported for Ca$_3$ReO$_5$Cl$_2$ \citen{Choi}, providing an ideal test for our calculations below.

The Raman spectra for Ca$_3$ReO$_5$Cl$_2$ displays a $180^\circ$ periodicity in $\phi$ for $\theta = 0$. This matches our predicted $\cos^2(\theta + \phi)$ intensity dependence. 
However, the reported Raman scattering for cross polarisation does not vanish in Ca$_3$ReO$_5$Cl$_2$. We atribute this to a non-magnetic or non-intrinsic contribution, and therefore compare the difference between the parallel and cross polarisation Raman scattering (blue curve in Fig. \ref{first_fig}) to our predictions.

We find excellent agreement between the  magnetic Raman intensity calculated for these interactions and that measured\cite{Choi} from Ca$_3$ReO$_5$Cl$_2$, Fig. \ref{first_fig}. For these calculations we take $J'/J = 0.3 $ and $J = 26.7~\text{cm}^{-1}$ from previous estimates\cite{hirai2019one, Choi} for Ca$_3$ReO$_5$Cl$_2$ and apply a Gaussian broadening of $6~\text{cm}^{-1}$ to all  theoretical spectra based on a conservative estimate of the width of the phonon peaks in Ca$_3$ReO$_5$Cl$_2$ at $4.3$~K\cite{Choi}. 
The theory has no free parameters. However, as the experiments are reported in arbitrary units only the lineshapes can be compared. 
The low-energy scattering is predicted correctly within the experimental noise, but we slightly underestimate the high energy tail; this is most likely due to higher order scattering, such as from states that include four spinons on a single chain \cite{caux2006four}, which we neglect. 

\begin{figure*}[t]
\centering
\includegraphics[width = 0.9\textwidth]{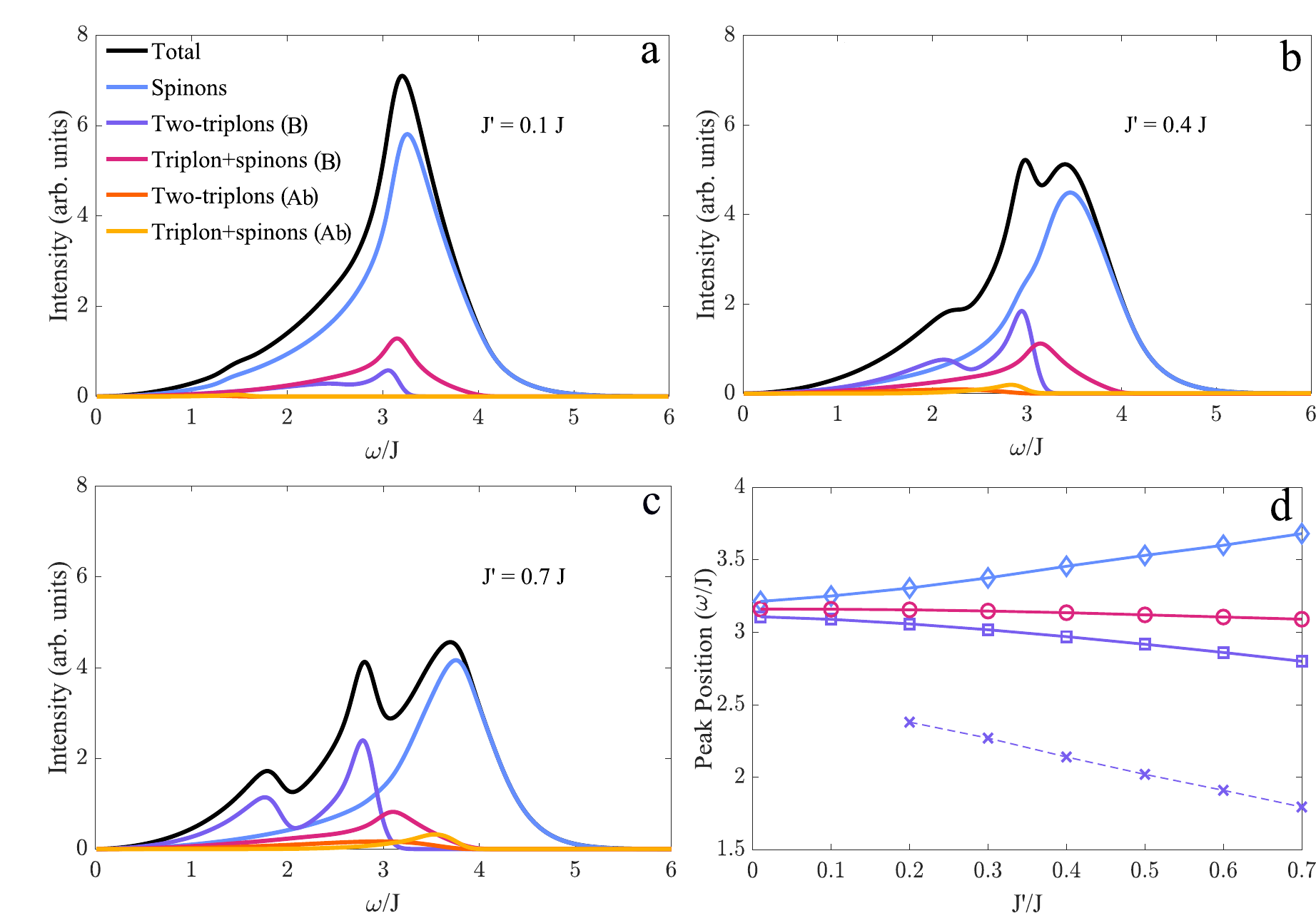}
\caption{\textbf{Increasing the interchain coupling of the anisotropic triangular antiferromagnet enhances the bound two-triplon contribution to the Raman intensity at the cost of reduced spinonic contribution.} (a) For small $J'/J$ all contributions to the magnetic Raman scattering are maximal near the triplon van Hove singularity at $\omega=\pi J$. (b,c) For larger interchain couplings the bound two-triplon and spinonic peaks separate and become more distinguishable. (d) Locations of the peaks with varying exchange anisotropy; lines between data points are to guide the eye. The two spinon peak (diamonds) is at $\omega \simeq \pi J + c_sJ'$, where $c_s\simeq 0.69$; the one triplon peak (circles) is at $\omega \simeq \pi J - c_1J'$, where $c_1\simeq 0.11$; the two triplon peak (squares) is at $\omega \simeq \pi J - c_2J'$, where $c_2\simeq 0.45$. A weaker third peak (crosses) emerges in the bound two-triplon signal for large $J'$, harbingering the onset of long-range order.
	The bound two-triplon peak is a generic feature of spinonic quasi-one-dimensional Raman scattering, and is therefore a clear spectral signature of dimensional reduction. }
\label{multi_fig}
\end{figure*}

The contributions to the  Raman intensity of Ca$_3$ReO$_5$Cl$_2$ from the different excitation classes is shown in Fig. \ref{second_fig}. The largest contribution  comes from spinonic excitations, which give rise to a large, asymmetrical peak in the intensity centred around $90~\text{cm}^{-1}$. The contributions from the bound two-triplon excitations and the bound-triplon and two-spinon excitations are comparable in size, both displaying peaks lower in energy than the spinonic peak. 
These two smaller peaks manifest a shoulder on the low energy side of the spinonic peak. 
The contributions involving antibound triplons are small compared to the experimental noise. 

In the q1D limit the magnetic Raman scattering displays two peaks near $\omega \simeq \pi J$, Figs. \ref{second_fig}, \ref{multi_fig}. A linear fit to our numerical data finds that the location of the peak arising from spinonic excitations as $\omega \simeq \pi J + c_s J'$, where $c_s \simeq 0.69$. Similarly, a linear fit to the peak associated to the bound two-triplon excitations yields $\omega = \pi J - c_2 J'$, where $c_2 \simeq 0.45$. Thus, the splitting between the peaks approximately scales as $\sim J'$.


The maximum in the two-triplon Raman scattering results from a van Hove singularity in the dispersion of the triplons. This occurs at the maximum energy of  bound triplons, which is of order $J'$ below the maximum of the lower bound of the two spinon continuum ($\omega \approx \pi J/2$)\cite{DCP}. 
Thus, this peak is expected for any system of weakly coupled Heisenberg chains, independent of the details of the lattice or interchain coupling. As such, the magnetic Raman peak at $\omega \simeq \pi J - c_2J'$ due to the bound two-triplon van Hove singularity is a smoking gun signature of fractionalisation in magnetic systems arising from q1D physics. We have explicitly confirmed that this peak remains on the rectangular lattice.

It is prudent to compare our results with the predictions of spin-wave theory for the anisotropic triangular lattice.\cite{perkins2013raman}  
For $J'/J\simeq0.34$  this predicts a peak at $\omega\simeq0.75J$ and a broad continuum around $\omega\simeq2J$. Both features are  well below $\pi J$ and therefore this calculation does not describe the Raman spectrum of Ca$_3$ReO$_5$Cl$_2$.\cite{Choi} Furthermore, these calculations predict the opposite dependence on the  polarisation angles to us. That is, they find $I_2(\omega)\gg I_1(\omega)$ for all $\omega$, yielding strong Raman scattering for cross polarisation and very weak (but non-vanishing)  scattering for parallel polarisation. This is the reverse to what is seen experimentally in Ca$_3$ReO$_5$Cl$_2$.\cite{Choi}

Only a few predictions have been made for Raman scattering from quantum spin liquids.\cite{PatrickLee,knolle2014raman,Cepas} While the details differ depending on the nature of the quantum spin liquids, each find broad continua with peaks at energies characteristic of the excitation spectrum of the relevant state. There is no reason to expect these to be near $\pi J$, and indeed none of the explicit calculations find peaks near $\pi J$. 

Therefore, the combination of a pair of peaks  around $\omega=\pi J$, split by $\sim J'$, with a $\cos^2(\theta+\phi)$ polarisation dependence of the Raman scattering, is a smoking gun for q1D physics. 


Our theory therefore makes clear predictions for a wide range of other materials. 
We predict that Raman scattering from Cs$_2$CuCl$_4$ will resemble that of Ca$_3$ReO$_5$Cl$_2$. 
Cu$_2$(OH)$_3$Br \cite{PhysRevMaterials.5.024407} and anhydrous alum KTi(SO$_4$)$_2$ \cite{Nilsen_2015} form approximately triangular lattices, with estimated interchain couplings on the order of $J' = 0.1 J$. We predict that the magnetic Raman signal in these materials will be  weak, as the signal strength scales as $J'^2$, with a broad Raman peak at approximately $\omega = 3.3J$ due largely to  spinons and a very weak shoulder below $\pi J$ due to the triplons, as shown in Fig. \ref{multi_fig}a. 
Sr$_3$ReO$_5$Cl$_2$, Ba$_3$ReO$_5$Cl$_2$, and $\kappa$-(ET)$_2$B(CN)$_4$ form anisotropic triangular $S=1/2$ antiferromagnets with $J'/J = 0.43$, 0.47, and 0.5 respectively\cite{hirai2020anisotropic,yoshida2015spin}, which may lead to an experimentally resolvable gap between the spinonic and bound two-triplon peaks in the Raman intensity, Fig. \ref{multi_fig}b,c.  
For these large values of $J'/J$ we find that  the bound two-triplon signal develops a second peak at  $\omega \approx 2J$ due to an enhancement of the transition rate at the zone boundary in the $\hat{x}$ direction (this becomes a weak shoulder at lower $J'$, e.g., Fig. \ref{second_fig}). Hence, for these large interchain coupling values we predict a triple peak structure to emerge. This is a  signature of the impending breakdown of the q1D picture.

More controversially, the quantum disordered state in EtMe$_3$Sb-[Pd(dmit)$_2$]$_2$ is usually atributated to a q2D QSL.\cite{Powell_RPP,savary2016quantum,zhou2017quantum,balents2010spin,knolle2019field} However, recent first principles calculations have suggested that a q1D anisotropic triangular lattice is a more appropriate model\cite{Kenny1,Kenny2,Ido2022}. 
If the prediction of Kenny \textit{et al}.\cite{Kenny1,Kenny2} that $J'/J\simeq0.32$ with $J\simeq$ 260-290$\text{~cm}^{-1}$ for EtMe$_3$Sb-[Pd(dmit)$_2$]$_2$ is correct, we would expect the magnetic Raman scattering to qualitatively resemble that of Ca$_3$ReO$_5$Cl$_2$ but with the broad peak occuring at around 800-920~cm$^{-1}$. Thus, magnetic Raman scattering could provide a direct test of the dimensionality of  EtMe$_3$Sb-[Pd(dmit)$_2$]$_2$.

Inelastic neutron scattering is a proven method for detecting spinons and triplons, as shown in Cs$_2$CuCl$_4$ and Ca$_3$ReO$_5$Cl$_2$, thereby demonstrating that these materials do not realise a quantum spin liquid state. We now have independent evidence due to magnetic Raman scattering, in conjunction with our theory, supporting this conclusion for the latter material. This positions magnetic Raman scattering as a powerful method for interrogating spin liquid candidates, particularly for materials where large single crystals are not available.

\section*{Methods}


We denote the one-dimensional Heisenberg antiferromagnetic groundstate on chain $y$ as $\ket{0}_y$ (hence $\ket{\text{GS}}_0 \equiv \otimes_y \ket{0}_y$), and a two-spinon excitation on chain $y$ with total momentum $k_x$, energy $\epsilon$, and total $z$-direction spin moment $S_z$ as $\ket{k_x, \epsilon, S_z}_y$. The selection rules of the Raman operator preclude singlet two-spinon states contributing to the exchange Raman intensity. 
The SU(2) symmetric representation of a singlet constructed from a pair of two-spinon excitations on different chains is denoted as
\begin{equation} \label{basis_state_1}
\begin{split}
\ket{\epsilon_1, \epsilon_2}_{\mathbf{k,p}} \equiv & \frac{-1}{ \sqrt{3}L_y}  \sum_{y_1, y_2, S_z} e^{i[\pi S_z + (k_y +p_y)\cdot y_1 + (k_y - p_y) \cdot y_2]}  \\
& \times \ket{k_x + p_x, \epsilon_1, S_z}_{y_1} \otimes \ket{k_x - p_x, \epsilon_2, -S_z}_{y_2} \\
&  \otimes_{y' \neq y_1 \neq y_2 \neq y'} \ket{0}_{y'}, 
\end{split}
\end{equation}
where $\epsilon_n$ ($\mathbf{k}_{n}$) is the energy (average momentum) of the $n$th pair of spinons, $\mathbf{k} \equiv \left(\mathbf{k}_{1} + \mathbf{k}_{2}\right)/2$ is the  average momentum of the spinon pairs, and $\mathbf{p} \equiv (p_x,p_y) \equiv \left(\mathbf{k}_{1} - \mathbf{k}_{2}\right)/2$ is the  relative momentum of the spinon pairs. 
To avoid double counting we restrict  $p_x$  to be strictly positive in the first Brillouin zone. 

As the photon momentum is always much smaller than the Fermi momentum, we are only concerned with the $\mathbf{k} = 0$ contributions to the Raman intensity.  
\begin{equation} \label{basis_state_2}
\begin{split}
 \hat{H}_0 \ket{\epsilon_1, \epsilon_2}_{\mathbf{0,p}} & = [ \epsilon_1({p}_x) + \epsilon_2(-{p}_x) + \omega_0 ] \ket{\epsilon_1, \epsilon_2}_{\mathbf{0,p}},
\end{split}
\end{equation}
where $\omega_0 = J (-\ln 2 + \frac{1}{4}) L_x L_y$ is the unperturbed groundstate energy of  $L_y$ Heisenberg chains, each with $L_x$  sites \cite{hulthen1938uber}. 

We  approximate the eigenstates by the states that contribute to the exchange Raman intensity at first  order in the interchain interaction and contain exactly two chains each with a two-spinon excitation: 
\begin{equation} \label{double_chain_eigenstate2}
\begin{split}
\ket{\lambda}   \simeq \frac{V}{(2\pi)^2} \int_{BZ} d\mathbf{p}\int d\epsilon_1 d\epsilon_2  D(p_x, \epsilon_1) D(p_x, \epsilon_2) & \\
 \times \Upsilon^{\omega_\lambda}_{\mathbf{p}} (\epsilon_1, \epsilon_2) \ket{\epsilon_{1}, \epsilon_{2}}_{0, \mathbf{ p}}&,
\end{split}
\end{equation}
where $V$ is the volume of the two-dimensional primitive cell, $\int_{BZ}$ denotes the integral over the first Brillouin zone, and 
$2 \pi D(p_x, \epsilon) / L_x$ is the density of states
of a two-spinon excitation with momentum $p_x$ and energy $\epsilon$, such that \cite{MullerBB}
\begin{equation} \label{DOS_early}
D(p_x, \epsilon) \equiv \frac{\Theta( \epsilon - \omega_L(p_x)) \Theta(\omega_U(p_x) -  \epsilon)}{\sqrt{\omega_U(p_x)^2 -  \epsilon^2}}, 
\end{equation}
where $\Theta(x)$ is the Heaviside step function and $\omega_L(p_x)$ and $\omega_U(p_x)$ are the lower and upper bounds of the two-spinon continuum given by the des Cloizeaux-Pearson relations \cite{DCP}. 

Hence, 
\begin{equation} \label{eng_expec}
\begin{split}
	\omega_\lambda =  & \omega_0 + \int_{BZ} d\mathbf{p} \int d\epsilon_1 d\epsilon_2 (\epsilon_1 + \epsilon_2) \\
& \times D( p_x, \epsilon_1)D( p_x, \epsilon_2) \abs{\Upsilon^{\omega_\lambda}_{\mathbf{p}} ( \epsilon_1, \epsilon_2)}^2 \\
& + \int_{BZ} d\mathbf{p} J' (\mathbf{p}) \int d\epsilon_{1}d\epsilon_{2}d\epsilon' D( p_x, \epsilon_1)D( p_x, \epsilon_2)\\
&\times D( p_x, \epsilon') A_{ p_x}(\epsilon')  \Upsilon^{*, \omega_\lambda}_{\mathbf{p}} ( \epsilon_1, \epsilon_2)  \\
& \times  \bigg[ A_{ p_x}(\epsilon_1) \Upsilon^{\omega_\lambda} _{\mathbf{p}} ( \epsilon', \epsilon_2)   + A_{ p_x}(\epsilon_2) \Upsilon^{\omega_\lambda} _{\mathbf{p}} ( \epsilon_1, \epsilon')  \bigg],
\end{split}
\end{equation}
where $J' (\mathbf{p}) = 4 \cos({p_x}/{2})\cos({\sqrt{3} p_y}/{2})$ is the Fourier transformed interchain coupling  and $A_{k}(\epsilon) \equiv (1/\sqrt{2}) \bra{0} S^{-}_{-k} \ket{k, \epsilon, +1}$ is the two-spinon spectral function, which is known exactly via the Bethe ansatz \cite{karbach1997two}. 
Thus, the $\omega_\lambda$ are given by the   solutions of the effective  Schr\"odinger equation 
\begin{equation} \label{eval_2}
\begin{split}
\omega_\lambda \Upsilon^{\omega_\lambda}_{\mathbf{p}} ( \epsilon_1, \epsilon_2) = & (\epsilon_1 + \epsilon_2 + \omega_0)  \Upsilon^{\omega_\lambda}_{\mathbf{p}} ( \epsilon_1, \epsilon_2) \\
&  +J' (\mathbf{p}) \int d\epsilon' D( p_x, \epsilon') A_{ p_x}(\epsilon')  \\
& \bigg[ A_{ p_x}(\epsilon_1) \Upsilon^{\omega_\lambda}_{\mathbf{p}} ( \epsilon', \epsilon_2)   + A_{ p_x}(\epsilon_2) \Upsilon ^{\omega_\lambda}_{\mathbf{p}} ( \epsilon_1, \epsilon')  \bigg].
\end{split}
\end{equation}

To evaluate the matrix element $_0\!\bra{\text{GS}} \hat{H}' \ket{\lambda}$ in equation \eqref{to_find_Gen} consider 
 \begin{equation} \label{zeroth_order3}
\begin{split}
\hat{H}' \ket{\lambda}  = & \sum_{k_x, y} J' \cos\left(\frac{k_x}{2}\right) \left( \mathbf{S}_{-k_x, y+1} + \mathbf{S}_{-k_x, y-1} \right) \cdot \mathbf{S}_{k_x, y} \\
& \times \int_{BZ} d\mathbf{p}\int d\epsilon_1 d\epsilon_2 D(p_x, \epsilon_1) D(p_x, \epsilon_2) \\
& \times \Upsilon^{\omega_\lambda}_{ \mathbf{p}} (\epsilon_1, \epsilon_2) \ket{\epsilon_{1}, \epsilon_{2}}_{0, \mathbf{ p}}. \\ 
\end{split}
\end{equation}
Noting that equation (\ref{eval_2}) is block diagonal in $\mathbf{p}$
, we find 
\begin{equation} \label{approx_elem}
\begin{split}
_0\!\bra{\text{GS}} \hat{H}' \ket{\lambda} 
= & -\sqrt{3} J' (\mathbf{p}) \int d\epsilon_1 d\epsilon_2  D(p_x, \epsilon_1) D(p_x, \epsilon_2) \\
& \times  A_{p_x}(\epsilon_1) A_{p_x}(\epsilon_2) \Upsilon^{\omega_\lambda}_{\mathbf{p}} (\epsilon_1, \epsilon_2). 
\end{split}
\end{equation}

Equations (\ref{eval_2}) and (\ref{approx_elem}) can be evaluated numerically by discretizing the two-spinon energy space $\epsilon$ into $Q$ points. The point spacing is governed by the discretization measure \cite{Kohno} 
    $\Delta D(p_x, \epsilon) = {\abs{p_x}}/{2Q}$, 
which guarantees that the distribution of points in $\epsilon$ becomes the two-spinon density of states in the limit $Q \rightarrow \infty$. Defining the discrete two-pairs wavefunction as $\Phi^{\omega_\lambda}_{\mathbf{p}} (\epsilon_1, \epsilon_2) \equiv ({\abs{p_x}}/{2Q}) \Upsilon^{\omega_\lambda}_{\mathbf{p}} (\epsilon_1, \epsilon_2)$, the effective Schr\"odinger equation (\ref{eval_2}) becomes
\begin{equation} \label{eigen_1_new}
\begin{split}
\omega_\lambda \Phi^{\omega_\lambda}_{\mathbf{p}} (\epsilon_1, \epsilon_2)  = & ~ (\epsilon_1 + \epsilon_2)  \Phi^{\omega_\lambda}_{\mathbf{p}} (\epsilon_1, \epsilon_2) \\
&  +J' (\mathbf{p}) \frac{\abs{p_x}}{2Q} \sum_{\epsilon_{3}} A_{ p_x}(\epsilon_3)  \\
&  \bigg[ A_{ p_x}(\epsilon_1) \Phi^{\omega_\lambda}_{\mathbf{p}} ( \epsilon_3, \epsilon_2)   + A_{ p_x}(\epsilon_2) \Phi^{\omega_\lambda}_{\mathbf{p}} ( \epsilon_1, \epsilon_3)  \bigg], \\
\end{split}
\end{equation}
which is a $Q^2 \cross Q^2$ eigenvalue problem. The $\mathbf{p}$ integrals were numerically evaluated on an evenly spaced $Q \cross Q$ grid. 
In all of our figures $Q = 150$.

\bibliographystyle{naturemag}
\bibliography{./MRS_triangular_paper_v5}

\section*{Acknowledgements}

We thank K. Y. Choi for sharing the previously published Raman data. This work was supported by the Australian Research Council (DP181006201) and MEXT Quantum Leap Flagship Program (MEXT Q-LEAP) Grant Number JPMXS0118069605.


\end{document}